\def\oneandahalfspace{\baselineskip=\normalbaselineskip
  \multiply\baselineskip by 3 \divide\baselineskip by 2}

\parskip=\medskipamount
\overfullrule=0pt 
\raggedbottom
\def\normalparindent{24pt}
\nopagenumbers
\footline={\ifnum\pageno=1{\hfil}\else{\hfil\rm\folio\hfil}\fi}
 
\def\beginlinemode{\endmode\begingroup\parskip=0pt 
                   \obeylines\def\\{\par}\def\endmode{\par\endgroup}}
\def\beginparmode{\endmode\begingroup \def\endmode{\par\endgroup}}
\let\endmode=\par
\def\raggedcenter{
                  \leftskip=2em plus 6em \rightskip=\leftskip
                  \parindent=0pt \parfillskip=0pt \spaceskip=.3333em 
                  \xspaceskip=.5em\pretolerance=9999 \tolerance=9999
                  \hyphenpenalty=9999 \exhyphenpenalty=9999 }
\def\\{\cr}
\let\rawfootnote=\footnote\def\footnote#1#2{{\parindent=0pt\parskip=0pt
        \rawfootnote{#1}{#2\hfill\vrule height 0pt depth 6pt width 0pt}}}
\def\title{\null\vskip 3pt plus 0.2fill\beginlinemode\raggedcenter\bf}
\def\author{\vskip 3pt plus 0.2fill \beginlinemode\raggedcenter}
\def\affil{\vskip 3pt plus 0.1fill\beginlinemode\raggedcenter\it}
\def\abstract{\vskip 3pt plus 0.3fill \beginparmode{\noindent  ABSTRACT:~}  }
   
\def\body{\beginparmode\parindent=\normalparindent}
\def\head#1{\par\goodbreak{\immediate\write16{#1} 
           {\noindent\bf #1}\par}\nobreak\nobreak}

\def\refto#1{$^{[#1]}$}
\def\ref#1{Ref.~#1}                     
\def\Ref#1{Ref.~#1}\def\cite#1{{#1}}\def\[#1]{[\cite{#1}]}
        
\def\(#1){(\call{#1})}
\def\call#1{{#1}}\def\taghead#1{{#1}}
\def\references{\head{REFERENCES}\beginparmode\frenchspacing\parskip=0pt}
\gdef\refis#1{\item{#1.\ }}
\def\endreferences{\body}
\def\endit{\endmode\vfill\supereject}\let\endpaper=\endit
\def\gsim{\mathrel{\raise.3ex\hbox{$>$\kern-.75em\lower1ex\hbox{$\sim$}}}}
\def\lsim{\mathrel{\raise.3ex\hbox{$<$\kern-.75em\lower1ex\hbox{$\sim$}}}}
\def\sla{\raise.15ex\hbox{$/$}\kern-.72em}
\def\iafedir{Instituto de Astronom\'\i a y F\'\i sica del Espacio\\Casilla de
          Correo 67 - Sucursal 28, 1428 Buenos Aires, Argentina}

\catcode`@=11
\newcount\r@fcount \r@fcount=0\newcount\r@fcurr
\immediate\newwrite\reffile\newif\ifr@ffile\r@ffilefalse
\def\w@rnwrite#1{\ifr@ffile\immediate\write\reffile{#1}\fi\message{#1}}
\def\writer@f#1>>{}
\def\referencefile{\r@ffiletrue\immediate\openout\reffile=\jobname.ref%
  \def\writer@f##1>>{\ifr@ffile\immediate\write\reffile%
    {\noexpand\refis{##1} = \csname r@fnum##1\endcsname = %
     \expandafter\expandafter\expandafter\strip@t\expandafter%
     \meaning\csname r@ftext\csname r@fnum##1\endcsname\endcsname}\fi}%
  \def\strip@t##1>>{}}

\def\citeall#1{\xdef#1##1{#1{\noexpand\cite{##1}}}}
\def\cite#1{\each@rg\citer@nge{#1}}
\def\each@rg#1#2{{\let\thecsname=#1\expandafter\first@rg#2,\end,}}
\def\first@rg#1,{\thecsname{#1}\apply@rg}       
\def\apply@rg#1,{\ifx\end#1\let\next=\relax%
\else,\thecsname{#1}\let\next=\apply@rg\fi\next}%
\def\citer@nge#1{\citedor@nge#1-\end-}  
\def\citer@ngeat#1\end-{#1}
\def\citedor@nge#1-#2-{\ifx\end#2\r@featspace#1 
  \else\citel@@p{#1}{#2}\citer@ngeat\fi}        
\def\citel@@p#1#2{\ifnum#1>#2{\errmessage{Reference range #1-#2\space is bad.}
    \errhelp{If you cite a series of references by the notation M-N, then M and
    N must be integers, and N must be greater than or equal to M.}}\else%
{\count0=#1\count1=#2\advance\count1 by1\relax\expandafter\r@fcite\the\count0,%
  \loop\advance\count0 by1\relax
    \ifnum\count0<\count1,\expandafter\r@fcite\the\count0,%
  \repeat}\fi}
\def\r@featspace#1#2 {\r@fcite#1#2,}    \def\r@fcite#1,{\ifuncit@d{#1}      
    \expandafter\gdef\csname r@ftext\number\r@fcount\endcsname%
    {\message{Reference #1 to be supplied.}\writer@f#1>>#1 to be supplied.\par
     }\fi\csname r@fnum#1\endcsname}
\def\ifuncit@d#1{\expandafter\ifx\csname r@fnum#1\endcsname\relax%
\global\advance\r@fcount by1%
\expandafter\xdef\csname r@fnum#1\endcsname{\number\r@fcount}}
\let\r@fis=\refis   \def\refis#1#2#3\par{\ifuncit@d{#1}%
    \w@rnwrite{Reference #1=\number\r@fcount\space is not cited up to now.}\fi%
  \expandafter\gdef\csname r@ftext\csname r@fnum#1\endcsname\endcsname%
  {\writer@f#1>>#2#3\par}}
\def\r@ferr{\endreferences\errmessage{I was expecting to see
\noexpand\endreferences before now;  I have inserted it here.}}
\let\r@ferences=\references
\def\references{\r@ferences\def\endmode{\r@ferr\par\endgroup}}
\let\endr@ferences=\endreferences
\def\endreferences{\r@fcurr=0{\loop\ifnum\r@fcurr<\r@fcount
    \advance\r@fcurr by 1\relax\expandafter\r@fis\expandafter{\number\r@fcurr}%
    \csname r@ftext\number\r@fcurr\endcsname%
  \repeat}\gdef\r@ferr{}\endr@ferences}
\let\r@fend=\endpaper\gdef\endpaper{\ifr@ffile
\immediate\write16{Cross References written on []\jobname.REF.}\fi\r@fend}
\catcode`@=12
\citeall\refto\citeall\ref\citeall\Ref 
\catcode`@=11
\newcount\tagnumber\tagnumber=0
\immediate\newwrite\eqnfile\newif\if@qnfile\@qnfilefalse
\def\write@qn#1{}\def\writenew@qn#1{}
\def\w@rnwrite#1{\write@qn{#1}\message{#1}}
\def\@rrwrite#1{\write@qn{#1}\errmessage{#1}}
\def\taghead#1{\gdef\t@ghead{#1}\global\tagnumber=0}
\def\t@ghead{}\expandafter\def\csname @qnnum-3\endcsname
  {{\t@ghead\advance\tagnumber by -3\relax\number\tagnumber}}
\expandafter\def\csname @qnnum-2\endcsname
  {{\t@ghead\advance\tagnumber by -2\relax\number\tagnumber}}
\expandafter\def\csname @qnnum-1\endcsname
  {{\t@ghead\advance\tagnumber by -1\relax\number\tagnumber}}
\expandafter\def\csname @qnnum0\endcsname
  {\t@ghead\number\tagnumber}
\expandafter\def\csname @qnnum+1\endcsname
  {{\t@ghead\advance\tagnumber by 1\relax\number\tagnumber}}
\expandafter\def\csname @qnnum+2\endcsname
  {{\t@ghead\advance\tagnumber by 2\relax\number\tagnumber}}
\expandafter\def\csname @qnnum+3\endcsname
  {{\t@ghead\advance\tagnumber by 3\relax\number\tagnumber}}
\def\equationfile{\@qnfiletrue\immediate\openout\eqnfile=\jobname.eqn%
  \def\write@qn##1{\if@qnfile\immediate\write\eqnfile{##1}\fi}
  \def\writenew@qn##1{\if@qnfile\immediate\write\eqnfile
    {\noexpand\tag{##1} = (\t@ghead\number\tagnumber)}\fi}}
\def\callall#1{\xdef#1##1{#1{\noexpand\call{##1}}}}
\def\call#1{\each@rg\callr@nge{#1}}
\def\each@rg#1#2{{\let\thecsname=#1\expandafter\first@rg#2,\end,}}
\def\first@rg#1,{\thecsname{#1}\apply@rg}
\def\apply@rg#1,{\ifx\end#1\let\next=\relax%
\else,\thecsname{#1}\let\next=\apply@rg\fi\next}
\def\callr@nge#1{\calldor@nge#1-\end-}\def\callr@ngeat#1\end-{#1}
\def\calldor@nge#1-#2-{\ifx\end#2\@qneatspace#1 %
  \else\calll@@p{#1}{#2}\callr@ngeat\fi}
\def\calll@@p#1#2{\ifnum#1>#2{\@rrwrite{Equation range #1-#2\space is bad.}
\errhelp{If you call a series of equations by the notation M-N, then M and
N must be integers, and N must be greater than or equal to M.}}\else%
{\count0=#1\count1=#2\advance\count1 by1\relax\expandafter\@qncall\the\count0,%
  \loop\advance\count0 by1\relax%
    \ifnum\count0<\count1,\expandafter\@qncall\the\count0,  \repeat}\fi}
\def\@qneatspace#1#2 {\@qncall#1#2,}
\def\@qncall#1,{\ifunc@lled{#1}{\def\next{#1}\ifx\next\empty\else
  \w@rnwrite{Equation number \noexpand\(>>#1<<) has not been defined yet.}
  >>#1<<\fi}\else\csname @qnnum#1\endcsname\fi}
\let\eqnono=\eqno\def\eqno(#1){\tag#1}\def\tag#1$${\eqnono(\displayt@g#1 )$$}
\def\aligntag#1\endaligntag  $${\gdef\tag##1\\{&(##1 )\cr}\eqalignno{#1\\}$$
  \gdef\tag##1$${\eqnono(\displayt@g##1 )$$}}
\def\eqalignno#1{\displ@y \tabskip\centering
  \halign to\displaywidth{\hfil$\displaystyle{##}$\tabskip\z@skip
    &$\displaystyle{{}##}$\hfil\tabskip\centering
    &\llap{$\displayt@gpar##$}\tabskip\z@skip\crcr
    #1\crcr}}
\def\displayt@gpar(#1){(\displayt@g#1 )}
\def\displayt@g#1 {\rm\ifunc@lled{#1}\global\advance\tagnumber by1
        {\def\next{#1}\ifx\next\empty\else\expandafter
        \xdef\csname @qnnum#1\endcsname{\t@ghead\number\tagnumber}\fi}%
  \writenew@qn{#1}\t@ghead\number\tagnumber\else
        {\edef\next{\t@ghead\number\tagnumber}%
        \expandafter\ifx\csname @qnnum#1\endcsname\next\else
        \w@rnwrite{Equation \noexpand\tag{#1} is a duplicate number.}\fi}%
  \csname @qnnum#1\endcsname\fi}
\def\ifunc@lled#1{\expandafter\ifx\csname @qnnum#1\endcsname\relax}
\let\@qnend=\end\gdef\end{\if@qnfile
\immediate\write16{Equation numbers written on []\jobname.EQN.}\fi\@qnend}
\catcode`@=12

\magnification=1200
\oneandahalfspace
\line{\hfill quant-ph/9607013}
\medskip
\title The  complete  solution  of  the  Hamilton-Jacobi  equation in 
{\centerline{Quantum Mechanics}}
\author Rafael Ferraro
\affil \iafedir
\affil Departamento de F{\'\i}sica, Facultad de Ciencias Exactas y Naturales
Universidad de Buenos Aires, Ciudad Universitaria, Pab. I
1428 Buenos Aires - Argentina
\bigskip
\bigskip
e-mail ferraro@iafe.uba.ar
\vskip 2cm
\line{PACS 03.65.Ca\hfill}
\abstract
An ordinary unambiguous integral representation  for the finite 
propagator of a quantum system is found by starting of a
privileged skeletonization of the functional action in phase space, provided by 
the complete solution of the Hamilton-Jacobi equation.
This representation allows to regard the propagator as the sum of the 
contributions coming from paths where the   momenta  generated by the complete 
solution of the Hamilton-Jacobi equation are conserved -as it does happen on 
the  classical trajectory-,  but are not  restricted  to  having  the classical
values associated with the boundary conditions for the original coordinates.
\eject
By taking Dirac's ideas \refto{d} into account, R.P.Feynman explained in 1948 
how Quantum Mechanics can be formulated from
principles that make contact with the variational principles
of Mechanics\refto{f}.   Feynman shown that Quantum Mechanics can be based
on the statement  that  the {\it propagator}, ie the probability amplitude of
finding the system in the state ${\bf q}''$ at $t''$, given that it was found
in ${\bf q}'$ at $t'$, can be obtained by means of the path integration:
$$<{\bf q}''\ t''|{\bf q}'\ t'>\ =\ \int
\ {\cal D}{\bf q}(t)\ \exp\left[{i\over \hbar}\ S[{\bf q}(t)]\right],
\eqno(1)$$
where $S[{\bf q}(t)]$ is the functional action of the system.
Since the  path  integral is a functional integration, one gives a meaning to
eq.\(1) by replacing each path  by a {\it skeletonized}  version  where the
path ${\bf q}(t)$ is represented  by interpolating points $({\bf q}_k, t_k)$,
$k = 0, 1,...,N$, ${\bf q}_0 = {\bf q}'$, ${\bf q}_N = {\bf q}''$.
Then the functional action  is replaced by a function $S(\{{\bf q}_k, t_k\})$,
and  the  functional integration reduces to  integrate  the  variables  ${\bf
q}_k$, $k = 1,...,N-1$.  Finally the  limit $\Delta t_k \equiv t_{k+1} - t_k$
$\rightarrow  0$ (ie, $N \rightarrow \infty$) is performed.    There  is  a
privileged recipe for the function $S(\{{\bf q}_k, t_k\})$ \refto{f}:
$$S(\{{\bf q}_k,t_k\})=\sum_{k=0}^{N-1}
S({\bf q}_{k+1} t_{k+1}\vert {\bf q}_k t_k),\eqno(2)$$ 
where $S({\bf q}_{k+1} t_{k+1}\vert {\bf q}_k t_k)$ is the Hamilton principal
function, ie the functional action evaluated on the classical path joining
$({\bf  q}_k,t_k)$ and $({\bf q}_{k+1},t_{k+1})$.   However  the  measure  
remains ambiguous in eq.\(1)\refto{dw,p,k}.  People have thought
that a path integration in phase space  could  remedy  this  problem  because
there is a privileged measure in phase space:   the  Liouville measure $d{\bf
q}\ d{\bf p}\  /(2\pi\hbar)^n$  ($n$  is  the  dimension of the configuration
space),  which  is  invariant  under  canonical transformations.
However there  was not found a privileged recipe to skeletonize the canonical
functional action 
$$S[{\bf q}(t),{\bf p}(t)]=\int_{t'}^{t''} \left({\bf p}(t)\cdot{\dot{\bf q}}(t)
-H({\bf q},{\bf p})\right)\ dt.\eqno(3)$$
In  \Ref{k,fe}    several  recipes  were  essayed  for    newtonian  and  
relativistic systems moving on a curved manifold.  The results showed that they
were equivalent  to  different  measures  in  eq.\(1), and different operator
orderings in the wave equation.  
\bigskip
In a general case,  $S[{\bf q}(t),{\bf p}(t)]$ should be replaced by a 
skeletonized action 
$$S(\{{\bf q}_k,{\bf p}_k,t_k\}) = \sum_{k=0}^{N-1} S({\bf q}_{k+1} t_{k+1}
\vert{\bf q}_k {\bf p}_k t_k)\eqno(3')$$ 
satisfying the following requirements \refto{k}:\par
\noindent{\it  i)} The points $({\bf q}_k,{\bf p}_k,t_k)$  are  interpolating
points  for the path ${\bf q}(t), {\bf p}(t)$.    Therefore  ${\bf  q}_0={\bf
q}'$, $t_0=t'$, and ${\bf q}_N={\bf q}''$, $t_N=t''$.\par
\noindent{\it ii)} The skeletonized action must be stationary on the points 
interpolating the classical path between  $({\bf  q}',t')$  and  
$({\bf  q}'',t'')$.\par
\noindent{\it  iii)}  When  $\Delta t_k\equiv (t_{k+1}-t_k)\rightarrow 0\ \ \
\forall k$,  the skeletonized action must go to the functional action for any
smooth path.
\par
\noindent{\it  iv)}  The    skeletonized  action  must  retain  the  symmetry
properties of the canonical functional action (for instance, invariance under
point  transformations,  ie  canonical  transformations    resulting  from  a
coordinate change in the configuration space).\par

Then the path integral
$$<{\bf q}''\ t''|{\bf q}'\ t'>\ =\ \int{\cal D}{\bf p}(t)
\  {\cal  D}{\bf    q}(t)\    \exp\left[{i\over\hbar}\    S[{\bf   q}(t),{\bf
p}(t)]\right],\eqno(4)$$
will be identified with
$$<{\bf q}''\ t''|{\bf q}'\ t'>\ =\ \displaystyle\lim_{\Delta t_k\to 0}\ 
\int {d{\bf p}_0\over(2\pi\hbar)^n}\left(\prod_{k=1}^{N-1} {d{\bf p}_k
\ d{\bf q}_k\over (2\pi\hbar)^n}\right)\ 
\exp\left[{i\over\hbar}\ \sum_{k=0}^{N-1} S({\bf q}_{k+1} t_{k+1}
\vert{\bf q}_{k} {\bf p}_k t_k)\right].\eqno(5)$$
We remark that ${\bf p}_0$ is integrated in eq.\(5), but ${\bf q}_0$ is not,
because ${\bf q}_0$  is the fixed boundary ${\bf q}'$.  The finite propagator
in eq.\(5) can be regarded as the composition of infinitesimal propagators:
$$\eqalign{<{\bf  q}''\  t''|{\bf q}'\  t'>\  =  \int  &<{\bf  q}''  t''|{\bf
q}_{N-1} t_{N-1}>\ d{\bf q}_{N-1}\ <{\bf q}_{N-1} t_{N-1}| .......\cr\cr
&.......|{\bf  q}_2 t_2>\ d{\bf q}_2\ <{\bf q}_2 t_2
|{\bf  q}_1 t_1>\ d{\bf q}_1\ <{\bf q}_1 t_1|{\bf q}' t'>,\cr}\eqno(5')$$
where each infinitesimal propagator is
$$<{\bf  q}''\  t''=t'+\epsilon|{\bf  q}'\  t'>  =  \int  {d{\bf  p}'\over
(2\pi\hbar)^n}\  \exp\left[{i\over\hbar}\  S({\bf  q}''\  t''\vert{\bf q}'\ 
{\bf p}'\  t')\right].\eqno(5'')$$
\bigskip
We are going to  show  that  the  complete  solution  of  the Hamilton-Jacobi
equation provides a privileged recipe to skeletonize the canonical action, in
the same way that the Hamilton  principal function  does in the configuration
space.    Let  be  $J({\bf    q},{\bf  P},t)$  a  complete  solution  of  the
Hamilton-Jacobi equation
$${\partial J\over\partial t} + H\left({\bf q}, {\partial J\over\partial {\bf
q}}\right) = 0,$$
where the ${\bf P}$'s are the $n$ integration  constants.    $J({\bf  q},{\bf
P},t)$  can  be  regarded  as  the generator of a  time  dependent  canonical
transformation $\{({\bf q}, {\bf p})\}\rightarrow$ $\{({\bf Q}, {\bf P})\}$
$${\bf p} = {\partial J\over\partial{\bf q}},\ \ \ \ \ \ \ \ \ \ \ 
{\bf Q} = {\partial J\over\partial{\bf P}}.\eqno(6)$$
The dynamical variables $\{({\bf Q}, {\bf P})\}$ result to be conserved on the
classical trajectory. We propose for the privileged skeletonization:
$$S({\bf q}_{k+1} t_{k+1}\vert{\bf q}_{k} {\bf p}_k t_k) =
J({\bf q}_{k+1},{\bf P}_k,t_{k+1}) - J({\bf q}_k,{\bf P}_k,t_k),\eqno(7)$$
where ${\bf P}_k = {\bf P}({\bf q}_k, {\bf p}_k, t_k)$.   The skeletonization
\(7) has a clear physical meaning in terms of the functional  action.   Since
$dJ = {\bf p}\cdot {\bf dq}$ $+ {\bf Q}\cdot {\bf dP} -  H  dt$, one realizes
that \(7) is the functional action evaluated on a path joining $({\bf q}_k,
t_k)$ with $({\bf q}_{k+1},t_{k+1})$  such  that  ${\bf  P}$ remains  constant
and equal to ${\bf P}_k$ along the path.  Although ${\bf P}$ does remain 
constant on the classical path, the paths associated with the skeletonization
\(7) are not classical in general, because the value ${\bf
P}_k$ is left free;  instead, on the classical path the value of ${\bf P}_k$ is
not arbitrary but is determined by the boundaries $({\bf q}_k,
t_k)$ and $({\bf q}_{k+1},t_{k+1})$.  In the spirit of \Ref{k}, the expression
\(7) will be   called {\it phase space principal function}.  
\bigskip
We will show that the skeletonization via the recipe \(7) fulfills the 
properties (ii)-(iv):\par
\medskip
\noindent{\it ii)} Let us consider ${\bf q}$
and ${\bf P}$ as independent variables, and begin by varying the skeletonized
action with respect to ${\bf P}_k$ .  It is a well know fact that the function 
$J({\bf q}'',{\bf P}',t'') - J({\bf q}',{\bf P}',t')$ evaluated at the point
${\bf P}'$ where it is stationary, is equal to the Hamilton principal function 
$S({\bf q}''\ t''\vert {\bf q}' t')$ \refto{l,ll}.  In fact, the condition
$${\partial\over\partial {\bf P}_k}J({\bf q}_{k+1},{\bf P}_k,t_{k+1}) -
{\partial\over\partial {\bf P}_k}J({\bf q}_{k},{\bf P}_k,t_{k})
=0,\ \ \ \ \ \ \ \ \ \forall k\eqno(8')$$
means that the ${\bf P}_k$'s are such that ${\bf Q}_{k+1} = {\bf  Q}_k$;  the
conservation  of  both  ${\bf  Q}$  and  ${\bf  P}$  implies that the path is
classical. Then  the stationary
value of the skeletonized action  \(7)  with  respect  to the variables ${\bf
P}_k$ coincides with the skeletonized action in the configuration space \(2).

By varying \(2) with respect to the ${\bf q}_k$'s, one gets the condition 
$${\partial\over\partial {\bf q}_k}S({\bf q}_{k+1} t_{k+1}\vert {\bf q}_k t_k)+
{\partial\over\partial {\bf q}_k}S({\bf q}_{k} t_{k}\vert {\bf q}_{k-1} t_{k-1})
=0,\ \ \ \ \ \ \ \ \ \forall k\eqno(8)$$
meaning that the ${\bf q}_k$'s are  such  that  the  final  momentum  of  the
classical path between $({\bf q}_{k-1},t_{k-1})$ and $({\bf q}_k,t_k)$,
matches the initial momentum of the classical path 
between $({\bf q}_k,t_k)$ and $({\bf q}_{k+1},t_{k+1})$. This 
continuity guarantees that the points $\{({\bf q}_k, {\bf p}_k)\}$ rendering
the skeletonized action \(7) stationary are interpolating points of the entire
classical path between $({\bf q}',t')$ and $({\bf q}'',t'')$.   

\noindent{\it iii)} For any smooth path, $\Delta{\bf q}_k \equiv$ 
${\bf  q}_{k+1} - {\bf q}_k$ goes to zero when $\Delta t_k \rightarrow 0$. Then
$$J({\bf    q}_{k+1},{\bf    P}_k,t_{k+1})    -  J({\bf  q}_k,{\bf  P}_k,t_k)
\longrightarrow  {\partial J\over\partial{\bf q}}\Big|_k\cdot \Delta{\bf q}_k +
{\partial J\over\partial t}\Big|_k\ \Delta t_k = {\bf p}_k\cdot
\Delta{\bf q}_k - H({\bf q}_k,{\bf p}_k) \Delta t_k.$$
Thus the skeletonized action \(7) goes to the functional action.

\noindent{\it iv)} It is obvious from the Hamilton-Jacobi equation 
that $J$  retains  the  invariances  of $H$:  if $H$ is invariant under point
transformations, then so are $J$ and the skeletonized action.

The infinitesimal quantum propagator of eq.\(5'') results in 
$$<{\bf q}''\ t''=t'+\epsilon|{\bf  q}'\  t'>  =  \int  {d{\bf  P}'\over
(2\pi\hbar)^n}\ 
\left\vert{\partial^2 J({\bf q}',{\bf P}',t')\over\partial{\bf q}'\partial{\bf
P}'}\right\vert\  \exp\left[{i\over\hbar}\ \left( J({\bf  q}'',{\bf  P}',
t'')  - J({\bf q}',{\bf P}',t')\right)\right],\eqno(8'')$$
where    $\left\vert    {\partial^2      J\over\partial{\bf    q}\partial{\bf
P}}\right\vert    =$    $\left\vert    {\partial{\bf      p}\over\partial{\bf
P}}\right\vert$ is the Jacobian for the  substitution  ${\bf p}\rightarrow{\bf
P}$.    

Note in  eqs.    \(5)  and  \(8'')  that  $<{\bf q}''\ t''|{\bf q}' t'>$ is a
bivaluated function which is scalar in ${\bf q}''$
but is a density of weight $1/2$ in ${\bf q}'$, 
because ${\bf p}'$ is integrated but ${\bf  p}''$  is
not.   These behaviors are compatible with the equation for the  propagation
of the wave function
$$\Psi({\bf q}'',t'')=\int d{\bf q}' <{\bf q}'' t''|{\bf q}'  t'>
\ \Psi({\bf q}',t'),\eqno(9')$$
if the  wave  function  $\Psi$  is going to be regarded as scalar.  An scalar
wave function compels to use an invariant measure $\mu({\bf q})\ d{\bf q}$ in
the inner product in the Hilbert space;  the density $\mu$ will be ultimately 
dictated by  the  result  of  the path integration \refto{fe}.  The different
behaviors of the  propagator \(5) under changes of ${\bf q}''$ and ${\bf q}'$
prevents the use of the notation 
${<{\bf q}''\ t''|{\bf q}'\ t'>^*}$ $= <{\bf  q}'\  t'|{\bf q}''\ t''>$.  
This lack of symmetry in  the  roles  played by ${\bf q}''$ and
${\bf q}'$ can be remedied in eq.\(8'') by splitting the Jacobian in
two factors depending on ${\bf q}''$ and ${\bf q}'$ respectively. 
Concretely, we propose to formulate the propagation of the wave function in 
Quantum Mechanics by {\it postulating} the following infinitesimal quantum
propagator: 
$$\eqalign{<{\bf q}''\  t''=t'+\epsilon|{\bf  q}'\  t'>  =  
\int  {d{\bf P}'\over(2\pi\hbar)^n}\ 
&\left\vert  {\partial^2 J({\bf q}'',{\bf P}', t'')\over\partial{\bf  q}''
\partial{\bf P}'}\right\vert^{1/2}\ 
\left\vert  {\partial^2 J({\bf q}',{\bf P}',t')\over\partial{\bf  q}'
\partial{\bf P}'}\right\vert^{1/2}\cr\cr 
&\exp\left[{i\over\hbar}\ \left( J({\bf  q}'',{\bf  P}', t'')  -
J({\bf q}',{\bf P}',t')\right)\right].\cr}\eqno(9)$$
This propagator does not depend  on  the  choice of the integration constants
${\bf P}$ in  the  Hamilton-Jacobi  equation,  because  it is invariant under
changes of ${\bf P}'$'s.  Since the
propagator \(9) is a density of weight $1/2$ in both arguments, then the wave
function is a density  of  weight  $1/2$;  therefore the inner product in the
Hilbert space is simply
$$(\Psi,\Phi) = \int d{\bf q} \ \Psi^*\ \Phi,\eqno(10)$$
no matter which generalized coordinates are used for describing the system.
\bigskip
Eq.\(9)  is  an  unambiguous  recipe for the propagator  that  is  privileged
because  of  its  direct  association  with the properties of  the  classical
system.
Moreover,  as  we  are  going  to  show,  the  composition  of  infinitesimal
propagators  leads  to  a finite quantum propagator of the {\it  same  form},
instead of a functional integration.  In fact,
let us consider the composition of infinitesimal propagators
$$\eqalign{<{\bf q}_3\ t_3&|{\bf q}_1\ t_1>\  =  \int \ d{\bf q}_2\  
<{\bf  q}_3\ t_3|{\bf  q}_2 t_2><{\bf q}_2 t_2|{\bf q}_1\ t_1> \cr\cr
=  &\int    d{\bf  q}_2\  {d{\bf  P}_2\over  (2\pi\hbar)^n}\  {d{\bf  P}_1\over
(2\pi\hbar)^n}\ 
\left\vert  {\partial^2 J\over\partial{\bf  q}_3
\partial{\bf P}_2}\right\vert^{1/2}\ 
\left\vert  {\partial^2 J\over\partial{\bf  q}_2\partial{\bf
P}_2}\right\vert^{1/2}\ 
\left\vert  {\partial^2 J\over\partial{\bf  q}_2
\partial{\bf P}_1}\right\vert^{1/2}\ 
\left\vert  {\partial^2 J\over\partial{\bf  q}_1\partial{\bf
P}_1}\right\vert^{1/2}\cr\cr 
&\exp\left[{i\over\hbar}\ \left( J({\bf  q}_3,{\bf  P}_2, t_3)  -
J({\bf q}_2,{\bf P}_2,t_2) + J({\bf  q}_2,{\bf  P}_1, t_2)  -
J({\bf q}_1,{\bf P}_1,t_1)\right)\right].\cr}\eqno(11)$$
In eq.\(11), the integral
$$\int \ d{\bf q}_2
\left\vert   {\partial^2    J({\bf    q}_2,{\bf    P}_2,t_2)\over\partial{\bf
q}_2\partial{\bf P}_2}\right\vert^{1/2}\ 
\left\vert  {\partial^2 J({\bf q}_2,{\bf P}_1,t_2)\over\partial{\bf  q}_2
\partial{\bf P}_1}\right\vert^{1/2}\ 
\exp\left[{i\over\hbar}\ \left(J({\bf q}_2,{\bf P}_1,t_2) - J({\bf q}_2,{\bf
P}_2, t_2)\right)\right],\eqno(12)$$
is a density of weight $1/2$ in ${\bf P}_1$ and ${\bf P}_2$.  A comparison with
eq.\(9) suggests that this  integral  is  equal  to $<{\bf P}_2|{\bf P}_1>$ $=
\delta({\bf P}_2-{\bf P}_1)$.  In order to confirm this suspect, one should
verify that $ J({\bf  q}_2,{\bf  P}_1,t_2)  -  J({\bf
q}_2,{\bf  P}_2, t_2)$ is a suitable skeletonized action   for  the  
${\bf P}$'s.  Since the variables  $\{({\bf  Q},{\bf P})\}$ generated
by the solution of the Hamilton-Jacobi equation are conserved, the functional
action which is stationary when the ${\bf P}$'s are  fixed at the extremes is
$$S[{\bf  Q},{\bf  P}] = \int_{t'}^{t''}\ {\bf P} \cdot\ \dot{\bf Q}\ dt\  -\
\left[{\bf  Q}\cdot{\bf  P}\right]_{t'}^{t''}  =  -\int_{t'}^{t''}\  {\bf  Q}
\cdot\ \dot{\bf P}\ dt,$$
while
$$  J({\bf  q}_2,{\bf  P}_1,t_2)  - J({\bf q}_2,{\bf P}_2, t_2)  =  {\partial
J\over\partial {\bf P}}\Big|_2\ \cdot\ ({\bf P}_1 - {\bf P}_2) = 
-{\bf Q}\cdot\Delta{\bf P}$$
for any smooth path.  Therefore we confirm that \(12) is the Dirac delta
$\delta({\bf P}_2 - {\bf P}_1)$, and the form \(5) of the quantum propagator 
will  remain unchanged even if the time interval is finite:
$$\eqalign{<{\bf q}''\ t''|{\bf q}'\ t'>\ =
\int {d{\bf P}\over (2\pi\hbar)^n}\ 
\left\vert  {\partial^2 J({\bf q}'',{\bf P},t'')\over\partial{\bf  q}''
\partial{\bf P}}\right\vert^{1/2}&\ 
\left\vert  {\partial^2 J({\bf q}',{\bf P},t')\over\partial{\bf  q}'
\partial{\bf P}}\right\vert^{1/2}\cr\cr
&\exp\left[{i\over\hbar}\ \left( J({\bf  q}'',{\bf  P},t'')  -
J({\bf q}',{\bf P},t')\right)\right].\cr}\eqno(13)$$
The finite quantum  propagator \(13) is an ordinary (not a functional) integral
which can be regarded as the superposition of contributions coming from paths 
joining  the  boundaries with arbitrary constant  values  of  the  classicaly
conserved dynamical variable ${\bf P}$.  The main contribution comes from the
classical path, where not only ${\bf P}$ but ${\bf  Q}$  is  conserved.    In
fact, the conservation of ${\bf Q}$ $= \partial J/\partial {\bf P}$ means that 
the classical path renders  the  phase  stationary.    The  knowledge  of the
classical  dynamics,  represented  by  the    complete    solution    of  the
Hamilton-Jacobi equation, determines without ambiguities the  propagation  of
the the wave function in Quantum Mechanics.

\vskip1cm
This work was partially supported by Consejo Nacional de Investigaciones
Cient{\'\i}ficas y T\'ecnicas (Argentina).
\bigskip
\references
\refis{l} C.Lanczos, {\it The Variational Principles  of  Mechanics}, Dover, New
York (1986).\par
\refis{ll}  L.D.Landau  and  E.M.Lifshitz, {\it Mechanics},  Pergamon  Press,
Oxford (1959).\par 
\refis{fe} R.Ferraro, Phys.Rev. D {\bf 45}, 1198 (1992).\par
\refis{f} R.P.Feynman, Rev.Mod.Phys. {\bf 20}, 367 (1948).\par
\refis{d} P.A.M.Dirac, Physik. Zeits. Sowjetunion {\bf 3}, 64 (1933).\par
\refis{dw} B.S.DeWitt, Rev.Mod.Phys. {\bf 29}, 377 (1957).\par
\refis{p} L.Parker, Phys.Rev. D {\bf 19}, 438 (1979).\par
\refis{k}K.Kucha{\v r}, J.Math.Phys. {\bf 24}, 2122 (1983).\par
\endreferences 
\end